\documentclass[12pt]{article}
\usepackage{fullpage}
\usepackage{amsmath,amsthm,amsfonts}

\newtheorem{theorem}{Theorem}%[section]
\newtheorem{lemma}[theorem]{Lemma}

\newtheorem{claim}[theorem]{Claim}

\newtheorem{definition}[theorem]{Definition}

\def\abs#1{\ensuremath{\left|{#1}\right|}}
\def\ceil#1{\ensuremath{\left\lceil{#1}\right\rceil}}
\def\floor#1{\ensuremath{\left\lfloor{#1}\right\rfloor}}

\def\BISC{\textsc{Cover}}

\begin{document}
%---
\title{On Covering a Graph Optimally with Induced~Subgraphs}
\author{Shripad Thite\thanks{%
Department of Mathematics and Computer Science, Technische
Universiteit Eindhoven, the Netherlands; Email: sthite@win.tue.nl}}
%\date{Last update: April 6, 2006}
\maketitle

\begin{abstract}
We consider the problem of covering a graph with a given number of
induced subgraphs so that the maximum number of vertices in each
subgraph is minimized.  We prove NP-completeness of the problem, prove
lower bounds, and give approximation algorithms for certain graph
classes.
\end{abstract}

Let $G=(V,E)$ be a graph. The \emph{order} of $G$ is the number
$\abs{V}$ of its vertices.  For an arbitrary subset of vertices $V'
\subseteq V$, the \emph{induced subgraph} denoted by $G[V']$ is the
subgraph of $G$ with vertex set $V'$ and all edges $e \in E$ such that
both endpoints of $e$ belong to $V'$.  In other words, $G[V']=(V',E')$
where $E' = \{ (u,v) \in E : {u,v \in V'} \}$. The \emph{union} of two
graphs $G_1(V_1,E_1)$ and $G_2(V_2,E_2)$ is the graph $G=(V_1 \cup
V_2, E_1 \cup E_2)$. We say that a graph $H$ \emph{covers} a graph $G$
if and only if $G$ is a subgraph of $H$.

We consider the following optimization problem: given a graph
$G=(V,E)$ and an integer $k \ge 0$, find $k$ induced subgraphs
$G[V_1]$, $G[V_2]$, $G[V_3]$, $\ldots$, $G[V_k]$ whose union covers
$G$ such that the maximum order of the induced subgraphs is minimized.
Thus, for every edge $(u,v)$ of $G$ we require that there exists an
$i$ in the range $1 \le i \le k$ such that both $u \in V_i$ and $v \in
V_i$; we wish to minimize $\max_{1 \le i \le k} \{\abs{V_i}\}$.  We
denote this problem by $\BISC(G,k)$.

Without loss of generality, we can assume that each $V_i$ has the same
cardinality, since we can add extra vertices to any subset smaller
than the largest without increasing the cost of the solution.

%% =================================================================
\paragraph{Motivation}

Suppose we have a parallel computer with $k$ processors, each with its
own local memory. The local memory of each processor is bounded, and
can store at most $M$ words.  We want to distribute $n$ items of data,
each occupying a single word in memory, among the processors so that
they can execute a certain computation in parallel.  An individual
step in the computation requires a processor to read a set of operands
from its memory, execute an operation, and write back the result again
to its local memory. Performing the operation requires that all
operands be present in the local memory of the processor.

We consider the case where the operations performed by the processor
are binary, i.e., each operation requires reading exactly two
operands. The computation is given as a graph $G=(V,E)$ with $n$
vertices where each vertex represents a data item and every edge of
the graph represents a dependency between two data items.  A processor
can execute the operation corresponding to an edge $(u,v)$ only if the
operands corresponding to both~$u$ and~$v$ are in the local memory of
the processor.

We wish to minimize the maximum required size of local memory among
all the processors so that every edge can be ``solved''. This requires
an assignment of data items or vertices to each of the $k$
processors. The subset of vertices assigned to processor $i$ is
$V_i$. We require that the induced subgraphs $G[V_i]$ together cover
the whole graph $G$. Minimizing the maximum local memory among the
processors is equivalent to minimizing the order of the largest
induced subgraph.

%% =================================================================
\paragraph{Related work}

Graph covering is a very well-studied problem---the online compendium
of NP-optimization problems~\cite{compendium}, for instance, lists
several NP-hard problems on partitioning and covering graphs.  Our
problem is different from each of the problems in the list and from
the many variants of graph covering, either because the constraints
are different (for instance, we do not require the covering subgraphs
to be connected or to be edge disjoint), or because the objective
function is different, or both.  To the best knowledge of the author,
no results on the particular problem we study in the current abstract
have been published yet.

%% *****************************************************************
%% *****************************************************************
%% *****************************************************************
%% *****************************************************************
\section{Complexity}
\label{sec:NPcomplete}

We show that deciding whether a forest can be covered optimally is
NP-complete.

The problem is clearly in NP.  We will show that it is NP-hard by
a reduction from \textsc{3-Partition} which is defined as follows:

\textsc{Given:} A set $A$ of $3m$ positive integer values $a_1$,
$a_2$, $\ldots$, $a_{3m}$ and a positive integer $S$ such that $S/4 <
a_i < S/2$ for all $i$ where $1 \le i \le 3m$ and such that
$\sum_{i=1}^{3m} a_i = mS$.\\
\textsc{Question:} Can $A$ be divided into $m$ disjoint subsets $B_1$,
$B_2$, $\ldots$, $B_m$ such that $\sum_{a_i \in B_j} a_i = S$ for all
$1 \le j \le m$?

Note that because $S/4 < a_i < S/2$ for all $i$, any solution that
answers the question in the affirmative must have $\abs{B_j} = 3$ for
all $j$.

\textsc{3-Partition} is known to be strongly
NP-complete~\cite{GareyJohnson}, i.e., it is NP-hard even when all
instances are encoded in unary.  We demonstrate a polynomial-time
reduction from an arbitrary instance of \textsc{3-Partition} to an
instance of $\BISC(G,k)$ that preserves ``yes'' and ``no'' answers.

The graph $G$ in the instance of $\BISC(G,k)$ will be a forest
of~$3m$ disjoint paths.  For each positive value~$a_i$ in the given
instance of \textsc{3-Partition}, construct a path~$P^{(i)}$
with~$a_i$ vertices and~$a_i-1$ edges.  Set~$k=m$.

If the original instance of \textsc{3-Partition} has a solution
consisting of $B_1$, $B_2$, $\ldots$, $B_m$, then we construct a
solution to the new instance of $\BISC(G,k)$ in which $G[V_j]$
is the union of the paths $P^{(i)}$ for all $i$ such that $a_i \in
B_j$.  Since the paths are disjoint, we have $\abs{V_j} = \sum_{a_i
\in B_j} a_i = S$ for all~$j$.  Hence, we obtain a solution to the
instance of $\BISC(G,k)$ of cost at most~$S$.

Next, consider a solution to the instance of $\BISC(G,k)$ in which
$\max_{1 \le j \le k} \abs{V_j} \le S$; hence $\sum_{1 \le j \le k}
\abs{V_j} \le k \max_{1 \le j \le k} \abs{V_j} = mS$.

The graph $G$ has $\sum_{1 \le i \le 3m} a_i = mS$ vertices.  Consider
the $mS \times k$ boolean incidence matrix that has a~$1$ entry if and
only if the corresponding vertex belongs to the corresponding subset.
The number of $1$'s in this incidence matrix is the sum of the
cardinalities of each~$V_j$.  Counting the number of $1$'s row-wise, we
see that the number of $1$'s is also equal to the sum over every
vertex $v$ of the number of subsets that contain~$v$.  Since each
vertex has positive degree and because every edge must be covered,
each vertex must belong to at least one subset.  Hence, it must be the
case that $\sum_{1 \le j \le k} \abs{V_j} \ge \sum_{1 \le i \le 3m}
a_i = mS$.  Therefore, we conclude that $\sum_{1 \le j \le k}
\abs{V_j} = mS$.  It follows that $\abs{V_j} = S$ for all~$j$ such
that $1 \le j \le m$.

Since~$G$ has~$mS$ vertices and $\sum_{1 \le j \le k} \abs{V_j} = mS$,
each vertex must belong to at most one subset.  Hence, each vertex
must belong to exactly one subset.  Since all the edges are covered,
any two adjacent vertices must belong to the same subset.  Therefore,
for each path $P^{(i)}$, all vertices of $P^{(i)}$ must belong to
exactly one~$V_j$.  Hence, we obtain a solution to the original
instance of \textsc{3-Partition} in which for all $i$ we have $a_i \in
B_j$ if and only if $P^{(i)}$ belongs to~$V_j$.  This is a valid
solution because for all $j$ we have $\sum_{a_i \in B_j} a_i =
\abs{V_j} = S$.

We have thus proved the following theorem.

\begin{theorem}
  $\BISC(G,k)$ is NP-hard even when~$G$ is a disjoint union
  of paths.
\end{theorem}

%% *****************************************************************
%% *****************************************************************
%% *****************************************************************
%% *****************************************************************
\section{Lower bounds}

In this section, we prove lower bounds on the size of an optimum
solution.

%----------------------------------
\subsection{A lower bound based on connectivity}

Clearly, $\max_{i}\ \abs{V_i} \geq \ceil{n/k}$. Let $\mathcal{I}$ be
the intersection graph of the $V_i$'s. If $G$ is connected, then
$\mathcal{I}$ must be connected, so it must have at least $k-1$
edges. Each edge of $\mathcal{I}$ corresponds to some vertex that
belongs to more than one subset, so $\sum_{i=1}^{k}\ \abs{V_i} \geq
n+k-1$. Since the maximum of a set is at least as large as its mean,
we have
\begin{equation}
\label{lb1}
\max_{i}\ \abs{V_i} \geq \ceil{\frac{n+k-1}{p}} = \ceil{\frac{n}{k}} + 1.
\end{equation}

Suppose $G$ is $\kappa$-connected. Let $N(V_i)$ denote the open
neighborhood of $V_i$; i.e. $N(V_i)$ consists of all vertices in the
complement of $V_i$ that are adjacent to some vertex in $V_i$.

Suppose $\max_i\ \{\abs{V_i}\} < n - \kappa$. We claim that
$\abs{N(V_i)} \geq \kappa$. This must be true because otherwise
removing the vertices in $N(V_i)$ would disconnect $V_i$ from the rest
of the graph (note that $\abs{\overline{V_i}} \geq \kappa$). Likewise,
$\abs{N(\overline{V_i})} \geq \kappa$.

For each subset $V_i$, let $W_i$ denote $\cup_{j \ne i} V_j$. Note
that $\overline{V_i} \subseteq W_i$.
$$\abs{V_i \cap N(\overline{V_i})} \geq \kappa \implies \abs{V_i \cap
W_i} \geq \kappa.$$
Hence, at least $\kappa$ vertices in each subset $V_i$ belong to more
than one subset.

%% .................................................................
\begin{claim}
\label{claim_sum}
$$\sum_{i=1}^{k}\ \abs{V_i} \geq n + \frac{\sum_{i=1}^{k}
\kappa_i}{2}$$
where $\kappa_i$ is the number of vertices of $V_i$ that belong to
more than one subset.
\end{claim}

\begin{proof}
The proof is by induction on the number of vertices, $n$. The base
case $n=0$ is trivial.  Remove a vertex $v$ of $G$. Suppose $v$
belongs to $s$ of the subsets. Therefore, in this smaller graph, from
the induction hypothesis, we have
$$\sum_{i=1}^{k}\ \abs{V_i} \geq n-1 + \frac{\left(\sum_{i=1}^{k}
\kappa_i\right) - s}{2}.$$

Hence, in the original graph,
\begin{align*}
\sum_{i=1}^{k}\ \abs{V_i} &\geq n + \frac{\left(\sum_{i=1}^{k}
\kappa_i\right) - s}{2} + s \\
  &\geq n + \frac{\sum_{i=1}^{k} \kappa_i}{2}.
\end{align*}
\end{proof}

From the previous argument, we have $\sum_{i=1}^{k} \kappa_i \geq k
\kappa$. Hence,
\begin{equation}
\label{lb2}
\max_{i}\ \abs{V_i} \geq \min\ \{n-\kappa, \ceil{\frac{n}{k}
+ \frac{\kappa}{2}}\}
\end{equation}
which is better than the lower bound of Equation~(\ref{lb1})
whenever $\kappa > 2$.

%--------------------------------------------------------
\subsection{A lower bound for dense graphs}
Let $\rho(m)$ be an upper bound on the number of edges in an induced
subgraph of $G$ of order $m$. The function $\rho$ is a measure of the
density of $G$. Any subset of $K$ or fewer vertices will cover at most
$\rho(K)$ edges. Since every edge in $G$ must be covered by the $k$
induced subgraphs, we must have
\[
\max_{1 \leq i \leq k}\ \{\abs{V_i}\} \geq \min\ \{ m : k \rho(m)
\geq e(G)\}.
\]

Since $\rho(m) \leq \binom{m}{2}$,
\[
\min\ \{ m : k \rho(m) \geq
e(G) \} \geq \min\ \{ m : k \binom{m}{2} \geq e(G) \};
\]
hence,
\begin{eqnarray}
\max_{1 \leq i \leq k}\ \{\abs{V_i}\}
\geq \frac{1}{2}\left(1 + \frac{\sqrt{8 e(G)+k}}{\sqrt{k}}\right)
>    \frac{1}{2}\left(1 + \sqrt{\frac{8 e(G)}{k}}\right)
>    \sqrt{\frac{2 e(G)}{k}}
\label{lb3}
\end{eqnarray}

For dense graphs where $e(G) = \Omega(n^2)$, equation (\ref{lb3})
gives us a lower bound of $\Omega(n \cdot k^{-1/2})$ which is better
than that of equation (\ref{lb2}). In particular, the bound of
$\Theta(n \cdot k^{-1/2})$ is tight when $G$ is a clique.

% -----------------------------------------------------------------
% -----------------------------------------------------------------
% -----------------------------------------------------------------
\subsection{Another lower bound}

Suppose~$V_1$, $V_2$, $\ldots$,~$V_k$ is a feasible solution (not
necessarily optimum) to~$\BISC(G,k)$; i.e.,~for every edge~$(u,v)
\in E$ there exists~$l$ such that~$\{u,v\} \subseteq V_l$.  Let~$S
\subseteq V$ be an arbitrary subset of vertices.  Let $N(S)$ denote the
neighborhood of the set~$S$, i.e.,~$N(S) = \{ v \in V \setminus S :
\exists u \in S, (u,v) \in E\}$.

Let~$C(S)$ denote~$\{ V_l : V_l \cap S \ne \emptyset, 1 \le l \le
k\}$.  We claim that~$S \cup N(S) \subseteq \bigcup_{V_l \in C(S)}
V_l$.  By definition of~$C(S)$, we have~$S \subseteq \bigcup_{V_l \in
C(S)} V_l$.  Let~$u \in S$ and~$v \in N(S)$ such that~$(u,v) \in E$.
Any subset~$V_l$ that covers the edge~$(u,v)$ must contain both~$u$
and~$v$ and, since~$V_l$ contains~$u \in S$, it must be the case
that~$V_l \in C(S)$.

Therefore,
\[
  \max_{V_l \in C(S)} \abs{V_l}
\ge
  \frac{\abs{S} + \abs{N(S)}}{\abs{C(S)}}
\]
In particular, we have shown that
\begin{equation}
  \max_{1 \le l \le k} \abs{V_l}
\ge
  \max_{S \subseteq V} (\abs{S} + \abs{N(S)})/k.
\label{eqn:lb-bushy}
\end{equation}

\bigskip

The question arises: how good are the lower bounds in this section?
The author suspects that they can be strengthened significantly, as
evidenced by the following lemma.

%% .................................................................
\begin{lemma}
  There exists an infinite family of trees such that, for every tree
  $T$ in the family with $n$ vertices, every optimum cover of $T$ with
  two induced subgraphs (i.e., $k=2$) must cost at least $\floor{n/2}
  + \Omega(\log n)$.
\end{lemma}

\begin{proof}
Construct a family of trees indexed by the integers inductively as
follows.  Each tree in the family will have a vertex designated as the
\emph{root}.  Let $T_0$ denote a tree with a single vertex which is
also the root.  For each $h \ge 1$, the tree $T_h$ consists of a new root
vertex plus three copies of $T_{h-1}$ such that the root of $T_h$ is
adjacent to the three roots of the copies of $T_{h-1}$.  It can be
easily verified that $T_h$ is a tree with $(3^h - 1)/2$ vertices.

For each tree $T_h$ in the above family, we apply the lower bound of
Equation~(\ref{eqn:lb-bushy}).  Let $V$ be the vertex set of
$T_h$. Let $S \subseteq V$ be an arbitrary subset of vertices such
that $\abs{S} = \abs{V}/2$. It can be shown that $\abs{N(S)} \ge
h-1$. Hence, the lemma follows by Equation~(\ref{eqn:lb-bushy}).
\end{proof}

%% *****************************************************************
%% *****************************************************************
%% *****************************************************************
%% *****************************************************************
\section{Approximation algorithms}

We turn our attention to specific graph classes and efficient
algorithms to approximate the optimum solution.

% =================================================================
% =================================================================
% =================================================================
% =================================================================
\subsection{Covering a caterpillar exactly}

A \emph{caterpillar} is a tree such that deleting all its leaves
causes a single path to remain.  Let $T$ be a caterpillar and let $V$
be the set of leaves of $T$; then, $T$ is a caterpillar if and only if
$T \setminus V$ is a single path $P$.

%% .................................................................
\begin{theorem}
  A caterpillar can be covered optimally by a greedy algorithm.
\end{theorem}

\begin{proof}
We show that a caterpillar $T$ can be covered optimally with exactly
$\ceil{n/k} + 1$ vertices in the induced subgraph of maximum order.
Order the vertices of $T$ in the following manner. Let $P$ be the
path that remains after deleting all leaves of $T$. Let $u$ be one of
the two endpoints of $P$. Choose $u$ to be the first vertex in the order,
followed by all leaves of $T$ adjacent to $u$ in arbitrary
order. Continue by ordering the vertices of $T \setminus u$ so that
they follow in the order.

Given the above vertex ordering, choose the prefix of the first
$\ceil{n/k}+1$ vertices as the set $V_1$. Remove the edges of $T$ in
the induced subgraph $G[V_1]$ and repeat the procedure on the
remaining graph until we have subsets $V_1$, $V_2$, $V_2$, $\ldots$,
$V_k$. The last subset $V_k$ may contain fewer vertices.

Note that no edge is covered by more than one induced subgraph. For
each $i$ in the range $1 \le i \le k-1$, the induced subgraph $G[V_i]$
is a subtree of $T$ with exactly $\ceil{n/k}+1$ vertices; hence,
$G[V_i]$ contains exactly $\ceil{n/k}$ edges. Therefore, $\bigcup_{1
\le i \le k-1} G[V_i]$ covers exactly $(k-1) \ceil{n/k}$ edges of
$T$. The remaining $(n-1) - (k-1)\ceil{n/k} \le n/k-1$ edges are easily
seen to be covered by $G[V_k]$ while ensuring that $\abs{V_k} \le n/k$.
\end{proof}

% =================================================================
% =================================================================
% =================================================================
% =================================================================
\subsection{Covering graphs of bounded degree}

Construct a vertex cover $\mathcal{C}$ of $G$ as follows: construct a
maximal matching and include both endpoints of each edge in the
matching. We get a vertex cover whose size is at most twice the
minimum possible. Let $\abs{\mathcal{C}} = c$.

Let $N[u]$ denote the closed neighborhood of a vertex $u$. (The closed
neighborhood of $u$ consists of $u$ and all vertices adjacent to
$u$. $N(u)$ denotes the \emph{open} neighborhood of $u$: $N(u) = N[u]
- \{u\}$.)  Then $\abs{N[u]} \leq \Delta + 1$, where $\Delta$ is the
maximum degree of $G$. Start with $c$ subsets of vertices, each
consisting of the closed neighborhood of a vertex in
$\mathcal{C}$. Clearly, every edge of $G$ has both endpoints in some
subset.

Assume that $c > k$.  Repeatedly merge the two smallest subsets, until
after $\floor{\lg (c/k)}$ steps we have only $k$ subsets. Each step at
most doubles the size of the largest subset. Therefore, at the end of
this process, $$\max_{i}\ \abs{V_i} \leq \frac{c}{k}\ (\Delta + 1).$$
The time taken by this process of merging is $O(\log (c/k)) = O(\log
(n/k))$.

Since the lower bound is $\ceil{n/k}$, this algorithm gives an
approximation ratio of
$$\frac{(\Delta + 1)\ c/k}{n/k} = \frac{c}{n}\ (\Delta + 1)
\leq \Delta + 1.$$
The total running time of the algorithm is easily seen to be linear in
the size of the graph.

% -----------------------------------------------------------------
% -----------------------------------------------------------------
% -----------------------------------------------------------------
\subsection{Covering $c$-inductive graphs}

An interesting class of graphs is the class of $c$-inductive (also
called $c$-degenerate) graphs.

\begin{definition}
  A graph~$G$ is \emph{$c$-inductive} if every subgraph of~$G$ has
  maximum degree at most~$c$.
\end{definition}

Equivalently, a graph~$G$ is $c$-inductive if it has a vertex~$u$ of
degree at most~$c$ such that $G \setminus u$ is $c$-inductive; the
empty graph is $c$-inductive by definition.

%% .................................................................
\begin{theorem}
  There exists an algorithm for $c$-inductive graphs with
  approximation ratio~$c+1$.
\end{theorem}

\begin{proof}
First, partition the $n$ vertices of $G$ into $k$ equitable subsets
$V_1$, $V_2$, $\ldots$, $V_k$, each of cardinality either
$\floor{n/k}$ or $\ceil{n/k}$.

Next, compute a $c$-inductive ordering of vertices as follows. Let
$v_1$ be a vertex of degree at most $c$ in $G$, let $v_2$ be a vertex
of degree at most~$c$ in $G-v_1$, and so on. In general, $v_i$ is a
vertex of degree at most $c$ in $G-\{v_1$, $v_2$, $\ldots$,
$v_{i-1}\}$; such a vertex must exist because $G$ is $c$-inductive.

Now, for each subset $V_i$ for $i=1,2,\ldots,k$, let $V'_i =
\emptyset$ initially. Consider each vertex $v_j \in V_i$ in the
inductive order restricted to vertices in~$V_i$. Include in $V'_i$ all
neighbors $v_l$ of $v_j$ with index greater than~$j$ such that $v_l
\in V \setminus (V_i \cup V'_i)$; due to the inductive ordering, there
are at most~$c$ neighbors of $v_j$ with index greater than~$j$. Thus,
$\abs{V'_i} \le c \abs{V_i} \le c \ceil{n/k}$. Finally, the desired
subsets of vertices are $V_i \cup V'_i$ for $1 \le i \le k$.

Suppose $(v_j, v_l)$ is an edge of $G$ with $j < l$ in the inductive
ordering. If both $v_j$ and $v_l$ belong to $V_i$ for some $i$, then
the edge $(v_j, v_l)$ is certainly covered. Otherwise, let $v_j \in
V_i$ and $v_l \in V_m$. When $v_j$ is encountered during the $i$th
stage $v_l$ is included in $V'_i$ if it is not in $V'_i$
already. Thus, every edge is covered when the algorithm terminates.

We have derived an upper bound of $(c+1) \ceil{n/k}$ on the
cardinality of $V_i \cup V'_i$ for every~$i$, which gives the
approximation ratio of the algorithm as~$c+1$.
\end{proof}

As a consequence, the above algorithm achieves in linear time a
$2$-approximation for forests (and trees), a $6$-approximation for
planar graphs, and a $3$-approximation for outerplanar graphs.

% -----------------------------------------------------------------
% -----------------------------------------------------------------
% -----------------------------------------------------------------
\subsection{Heuristic for graph classes with separator theorems}

A separator theorem for a class of graphs $\mathcal{G}$ is a theorem
of the following form~\cite{lipton79:separator}:
\begin{quote}
There exist constants $\alpha < 1$ and $\beta > 0$ such that if $G$ is
any $n$-vertex graph in $\mathcal{G}$, then the vertices of $G$ can be
partitioned into three subsets $A$, $B$, and $C$ such that no edge
joins a vertex in $A$ with a vertex in $B$, neither $A$ nor $B$
contains more than $\alpha n$ vertices, and $C$ contains at most
$\beta f(n)$ vertices.
\end{quote}
Such a subset $C$ is said to be an $(\alpha, \beta f(n))$-separator of
$G$.

A natural recursive algorithm for covering a graph $G \in \mathcal{G}$
with $k$ induced subgraphs is the following. Find subsets of vertices
$A$, $B$, and $C$ as above. Without loss of generality, assume that
$A$ has no more vertices than $B$. Recursively construct a cover with
$\floor{k/2}$ induced subgraphs of $G[A \cup C]$ and a cover with
$\ceil{k/2}$ induced subgraphs of $G[B \cup C]$.  The recursion
terminates when $k=1$ with the trivial solution.  Since $C$ is a
separator, every edge of $G$ belongs to either $G[A \cup C]$ or $G[B
\cup C]$; hence, we indeed obtain a cover.

The solution obtained by the above recursive algorithm is close to
optimal if $f(n) = o(n)$, i.e., if every graph in the class
$\mathcal{G}$ has a separator of sublinear order.

%% *****************************************************************
%% *****************************************************************
%% *****************************************************************
%% *****************************************************************
\section{A dual problem}

A natural dual problem is to cover a given graph with as few induced
subgraphs as possible, each with a fixed maximum number of vertices.
Given a graph $G = (V,E)$ and an integer $m$, cover $G$ with the
minimum number of induced subgraphs $G[V_i]$, $G[V_2]$, $\ldots$,
$G[V_p]$, such that $\abs{V_i} \leq m$ for all $i$.  Here, the problem
is to minimize the number of processors, each with a fixed amount of
local memory, to cover the given computation graph.

The dual problem is also NP-complete by the same proof as for the
primal; see Section~\ref{sec:NPcomplete}.

%% *****************************************************************
%% *****************************************************************
%% *****************************************************************
%% *****************************************************************
\section{Extension to covering hypergraphs}

The problem can be generalized to the case of covering a hypergraph.
The computation can be modeled by a hypergraph $\mathcal{H}$ with
vertex set $[n]$. Each edge of the hypergraph is a set of data items
that are operands of any single operation and are therefore required
to be stored together in some processor's memory. The problem is to
assign the vertices of $\mathcal{H}$ to $p$ subsets $V_1$, $V_2$,
$\ldots$, $V_p$ such that $\abs{V_i} \leq K$ and every edge of
$\mathcal{H}$ belongs to at least one of the subgraphs of
$\mathcal{H}$ induced by $V_1$, $V_2$, $\ldots$, $V_p$. In other
words, we need to cover $\mathcal{H}$ with $p$ induced subgraphs
$\mathcal{H}[V_1]$, $\mathcal{H}[V_2]$, $\ldots$, $\mathcal{H}[V_p]$
such that the order of each subgraph is at most $K$.

Note that, in general, a single vertex may belong to more than one
subset. Unlike in the graph case, a single hyperedge $e$ can be
covered more than once but only if there exists some other hyperedge
$f$ such that $e \subset f$, and $e$ and $f$ are covered by different
subgraphs. On the other hand, we can assume without loss of generality
that no hyperedge is contained in another.

%% *****************************************************************
%% *****************************************************************
%% *****************************************************************
%% *****************************************************************
\subsection*{Acknowledgments}

The author thanks Shang-Hua Teng for posing the original problem
several years ago. Thanks also to Jeff Erickson, Douglas West, Mitch
Harris, Sariel Har-Peled, and Alper \"Ung\"or for various
discussions.  Almost all the results mentioned in this abstract were
obtained when the author was a Ph.D.\@ student in the Computer Science
department at the University of Illinois at Urbana-Champaign.

%GATHER{D:\research\graphs\bisc\bisc.bib}
%\bibliographystyle{alpha}
%\bibliography{../bisc}

%---
\end{document}